# Electromagnetic Retarded Interaction and Symmetry Violation of Time Reversal in Light's High Order Stimulated Radiation and Absorption Processes

## Mei Xiaochun

(Institute of Theoretical Physics in Fuzhou, China, E-mial:mxc001@163.com)

**Abstract** It is proved that when the retarded effect (or multiple moment effect) of radiation fields is taken into account, the light's high order stimulated radiation and stimulated absorption probabilities are not the same so that time reversal symmetry would be violated, though the total Hamiltonian of electromagnetic interaction is still unchanged under time reversal. The reason to cause time reversal symmetry violation is that certain filial or partial transitive processes of bounding state atoms are forbidden or can't be achieved actually due to the law of energy conservation, the asymmetric actions of effective transition operators before and after time reversal, as well as the special states of atoms themselves. These restrictions would cause the symmetry violation of time reversal of other filial or partial transition processes which can be actualized really. The symmetry violation is also relative to the initial state's asymmetries of bounding atoms before and after time reversal. For the electromagnetic interaction between non-bounding state's atoms and radiation field, there is no this kind of symmetry violation of time reversal. In this way, the current formula of light's stimulated radiation and absorption parameters with time reversal symmetry should be revised. The influence of time reversal symmetry violation on the foundational theory of laser is also discussed and the phenomena of non-population inversion and non-radiation transition can also be explained well. In this way, a more reliable foundation can be established for the theories of laser and nonlinear optics in which non-equilibrium processes are involved.

## 1. Introduction

Einstein put forward the theory of light's stimulated radiation and absorption in 1917 in order to explain the Planck blackbody radiation formula based on equilibrium theory. According to the Einstein's theory, the parameters of stimulated radiation and absorption are equal to each other with $B_{ml} = B_{lm}$. The same result can also be obtained by means of the calculation of quantum mechanics for the first order process under dipole approximation without considering the retarded interaction (or multiple moment effect) of radiation fields [1]. Because light's stimulated radiation process can be regarded as the time reversal of stimulated absorption process, the result means that light's stimulated radiation and absorption processes are with time reversal symmetry.

Nonlinear optics was advanced in the 1960s. Also by the dipole approximation without considering the retarded interaction of radiation fields, nonlinear susceptibilities in nonlinear optics are still invariable under time reversal [2]. So the processes of light's radiation and absorption as well as nonlinear optics are considered with time reversal symmetry at present. In fact, it is a common and wide accepted idea at present that all micro-processes controlled by electromagnetic interaction are symmetrical under time reversal, for the motion equations of quantum mechanics and the Hamiltonian of electromagnetic interaction are unchanged under time reversal.

However, most processes relative to laser and nonlinear optics are actually high non-equilibrium ones. As we known that time reversal symmetry would generally be violated in non-equilibrium processes. It is proved below that after the retarded effect of radiation fields is taken into account, time reversal symmetry would be violated in light's high order stimulated radiation and absorption processes with $B_{ml} \neq B_{lm}$, though the Hamiltonian of electromagnetic interaction is still unchanged under time reversal. The main reason to cause time reversal symmetry violation is that certain filial or partial transition processes of bounding state atoms are forbidden or can't be achieved due to the law of energy conservation, the asymmetric actions of effective transition operators before and after time reversal, as well as the special states of atoms themselves. These restrictions would cause the symmetry violation of time reversal of other filial or partial transition processes which can be actualized really. These realizable filial or partial processes are just the practically observable physical processes which violate time reversal symmetry generally. The symmetry violation is also relative to the initial state's asymmetries of bounding atoms before and after time reversal. For the electromagnetic interaction between non-bounding atoms and radiation fields, there is no this kind of symmetry violation of time reversal. In fact, a great number of experiments have shown that the production processes of laser and most of non-linear optical processes, just as the optical processes of sum frequency, double frequency and different frequency, double stable states [3], self-focusing and self-defocusing [4], echo phenomena [5], as well as optical self-transparence and self observations [6] and so on, are obviously violate the time reversal symmetry. Only because the current theory does not think that time reversal symmetry violation would exists in micro- processes, physicists look at but can not see them. In this way, the current formula of light's stimulated radiation and absorption parameters with time reversal symmetry should be revised. A more reliable foundation can be established for the theories of laser and nonlinear optics in which non-equilibrium processes are involved.

## 2. The transition probability of the first order process

For simplification, we consider an atom with an electron in its external layer. Electron's mass is $\mu$, charge is $q$. When there is no external interaction, the Hamiltonian and wave function of the electron are individually

$$\hat{H}_0 = -\frac{\hbar^2}{2m}\nabla^2 + \hat{U}(r) \qquad |\psi_0\rangle = \sum_n e^{-\frac{1}{\hbar}E_n t}|n\rangle \qquad (1)$$

After external electromagnetic field is introduced, the interaction Hamiltonian is

$$\hat{H}' = -\frac{q}{c\mu}\hat{A}\cdot\hat{p} + \frac{q^2}{2c^2\mu}\hat{A}^2 + q\varphi \qquad (2)$$

When the charge and current densities of radiation field are zero, we can take the gauge condition $\nabla\cdot\hat{A} = 0$ and $\varphi = 0$ and write $\hat{H}' = \hat{H}'_1 + \hat{H}'_2$ with

$$\hat{H}'_1 = -\frac{q}{c\mu}\hat{A}\cdot\hat{p} \qquad \hat{H}'_2 = \frac{q^2}{2c^2\mu}\hat{A}^2 \qquad (3)$$

Where $\hat{H}'_1$ is with the order $v/c$ and $\hat{H}'_2$ is with the order $v^2/c^2$. In the current discussion for light's stimulated radiation and absorption theory, $\hat{H}'_2$ is neglected generally. Because $\hat{H}'_2$ has the same order of magnitude as the second order effects of nonlinear optics, it is remained in the paper. Suppose that electromagnetic wave propagates along $\vec{k}$ direction. Electric field strength is $\vec{E} = \vec{E}_0 \sin(\omega t - \vec{k} \cdot \vec{R})$. Here $\vec{R}$ is a direction vector pointing from wave source to the observation point. Both $\hat{H}'_1$ and $\hat{H}'_2$ can also be written as

$$\hat{H}'_1 = -\frac{q\vec{E}_0}{2\omega\mu} \cdot \left[ e^{i(\omega t - \vec{k} \cdot \vec{R})} + e^{-i(\omega t - \vec{k} \cdot \vec{R})} \right] \hat{p} \tag{4}$$

$$\hat{H}'_2 = \frac{q^2\vec{E}_0^2}{2c^2k^2\mu} \left[ e^{i(\omega t - \vec{k} \cdot \vec{R})} + e^{-i(\omega t - \vec{k} \cdot \vec{R})} \right]^2 = \frac{q^2\vec{E}_0^2}{2\omega^2\mu} \left[ e^{i2(\omega t - \vec{k} \cdot \vec{R})} + e^{-i2(\omega t - \vec{k} \cdot \vec{R})} + 2 \right] \tag{5}$$

We can write $\hat{H}'_1 = \hat{F}_1 e^{i\omega t} + \hat{F}_1^+ e^{-i\omega t}$ in which

$$\hat{F}_1 = -\frac{q\vec{E}_0}{2\omega\mu} \cdot e^{-i\vec{k} \cdot \vec{R}} \hat{p} \qquad\qquad \hat{F}_1^+ = -\frac{q\vec{E}_0}{2\omega\mu} \cdot e^{i\vec{k} \cdot \vec{R}} \hat{p} \tag{6}$$

Because we always have $\vec{k} \perp \vec{E}_0$ for electromagnetic wave, we have the commutation relation $[\vec{k} \cdot \vec{R}, \vec{E}_0 \cdot \hat{p}] = i\hbar \vec{k} \cdot \vec{E}_0 = 0$. So it can be proved that $\vec{E}_0 \cdot \hat{p}$ and $\exp(i\vec{k} \cdot \vec{R})$ are also commutative. In this way, Eq.(6) can also be written as

$$\hat{F}_1 = -\frac{q\vec{E}_0}{2\omega\mu} \cdot \hat{p} e^{-i\vec{k} \cdot \vec{R}} \qquad\qquad \hat{F}_1^+ = -\frac{q\vec{E}_0}{2\omega\mu} \cdot \hat{p} e^{i\vec{k} \cdot \vec{R}} \tag{7}$$

By using perturbation method in quantum mechanics to regard $\hat{H}'$ as perturbation, we write the motion equation and wave function of system as

$$i\hbar \frac{\partial}{\partial t} |\psi\rangle = (\hat{H}_0 + \hat{H}') |\psi\rangle \qquad\qquad |\psi\rangle = \sum_m a_m(t) e^{-\frac{i}{\hbar} E_m t} |n\rangle \tag{8}$$

Let $a_m(t) = a_m^{(0)}(t) + a_m^{(1)}(t) + a_m^{(2)}(t) + \cdots$, $E_m - E_n = \hbar\omega_{mn}$, substitute them into the motion equation, we can get

$$i\hbar \frac{d}{dt} \left[ a_m^{(0)}(t) + a_m^{(1)}(t) + a_m^{(2)}(t) + \cdots \right] = \sum_n \left( \hat{H}'_{1mn} + \hat{H}'_{2mn} \right) \left[ a_n^{(0)}(t) + a_n^{(1)}(t) + a_n^{(2)}(t) + \cdots \right] e^{i\omega_{mn} t} \tag{9}$$

The items with same order on the two sides of the equation are taken to be equal to each other. The first four equations are

$$i\hbar \frac{d}{dt} a_m^{(0)}(t) = 0 \qquad\qquad i\hbar \frac{d}{dt} a_m^{(1)}(t) = \sum_n \hat{H}'_{1mn} a_n^{(0)}(t) e^{i\omega_{mn} t} \tag{10}$$

$$i\hbar \frac{d}{dt} a_m^{(2)}(t) = \sum_n \hat{H}'_{2mn} a_n^{(0)}(t) e^{i\omega_{mn} t} + \sum_n \hat{H}'_{1mn} a_n^{(1)}(t) e^{i\omega_{mn} t} \tag{11}$$

$$i\hbar \frac{d}{dt} a_m^{(3)}(t) = \sum_n \hat{H}'_{2mn} a_n^{(1)}(t) e^{i\omega_{mn}t} + \sum_n \hat{H}'_{1mn} a_n^{(2)}(t) e^{i\omega_{mn}t} \qquad (12)$$

Suppose that an electron is in the initial state $|l\rangle$ with energy $E_l$ at time $t=0$, then the electron transits into the final state $|m\rangle$ with energy $E_m$ at time $t$, we have $a_m^{(0)}(t) = \delta_{ml}$ from the first formula of Eq.(10). Put it into the second formula, the probability amplitude of the first order process is

$$a_m^{(1)}(t) = \frac{1}{i\hbar} \int_0^t \hat{H}'_{1ml} e^{i\omega_{ml}t} dt = -\frac{\hat{F}_{1ml}\left[e^{i(\omega+\omega_{ml})t} - 1\right]}{\hbar(\omega+\omega_{ml})} + \frac{\hat{F}_{1ml}^+ \left[e^{-i(\omega-\omega_{ml})t} - 1\right]}{\hbar(\omega-\omega_{ml})} \qquad (13)$$

Where $\hat{F}_{1ml} = \langle m|\hat{F}_1|l\rangle$, $\hat{F}_{1ml}^+ = \langle m|\hat{F}_1^+|l\rangle$. The formula represents the probability amplitude of an electron transiting from the initial state $|l\rangle$ into the final state $|m\rangle$. In current theory, the so-called rotation wave approximation is used, i.e., only the first item is considered when $\omega = -\omega_{ml}$ and the second one is considered when $\omega = \omega_{ml}$ in Eq.(13). But up to now we have not decided which one is the state of high-energy level and which one is on the state of low-energy level. In fact, electron can either transit into higher-energy state $|m\rangle$ from low-energy state $|l\rangle$ by absorbing a photon, or transit into low-energy state $|m\rangle$ from high-energy state $|l\rangle$ by emitting a photon. Because photon's energy is always positive, by the law of energy conservation, it is proper for us to think that the condition $\omega = \omega_{ml}$ corresponds to the situation with $E_m > E_l$, in which an electron transits into the high-energy final state $|m\rangle$ from the low-energy initial state $|l\rangle$ by absorbing a photon with energy $\hbar\omega_{ml} = E_m - E_l > 0$. This is just the simulated absorption process with the transition probability in unit time

$$W_{\omega=\omega_{ml}}^{(1)} = \frac{2\pi}{\hbar^2} \left|\hat{F}_{1ml}^+\right|^2 \delta(\omega - \omega_{ml}) \qquad (14)$$

Therefore, the condition $\omega = -\omega_{ml}$ corresponds to the situation with $E_m < E_l$, in which an electron transits from the high-energy initial state $|l\rangle$ into the low-energy final state $|m\rangle$ by emitting a photon with energy $-\hbar\omega_{ml} = \hbar\omega_{lm} = E_l - E_m > 0$. This is just the simulated radiation process with the transition probability in unit time

$$W_{\omega=-\omega_{ml}}^{(1)} = \frac{2\pi}{\hbar^2} \left|\hat{F}_{1ml}\right|^2 \delta(\omega + \omega_{lm}) \qquad (15)$$

So $W_{\omega=\omega_{ml}}^{(1)}$ and $W_{\omega=-\omega_{ml}}^{(1)}$ represent the different physical processes. It is necessary for us to distinguish the physical meanings of $W_{\omega=\omega_{ml}}^{(1)}$ and $W_{\omega=-\omega_{ml}}^{(1)}$ clearly for the following discussion. As shown in Fig.1, we image a system with three energy levels: medium energy level $E_l$, high energy $E_m(up)$ and low energy $E_m(down)$. The difference of energy levels between $E_l$ and $E_m(up)$ is the same as that between $E_m(down)$ and $E_l$. Suppose that the electron is in medium energy level at beginning. Stimulated by radiation field, the electron can either transit up into high-energy level or down into low energy level. In this case, $W_{\omega=\omega_{ml}}^{(1)}$ represents the probability the electron transits up into high-energy level and $W_{\omega=-\omega_{ml}}^{(1)}$ represents the probability the electron transits down into low-energy level.

For visible light with wavelength $\lambda \sim 10^{-7} m$ and common atoms with radius $R \sim 10^{-10} m$, we have $\vec{k} \cdot \vec{R} \sim 10^{-3} \ll 1$. So in the current theory, dipolar approximation $\exp i\vec{k}\cdot\vec{R} \sim 1$ is taken into account. However, it should be noted that the ratio of magnitude between the first order processes and the second order processes in nonlinear optics is just about $10^{-3}$. Meanwhile, for the interaction between external fields and electrons in atoms, such as the situations of laser and nonlinear optics, we have $R \approx 0.1 \sim 1m$ so that $\vec{k} \cdot \vec{R} = 10^6 \sim 10^7$ with the macro-order of magnitude. In fact, factor $\vec{k} \cdot \vec{R}$ represents the retarded interaction of electromagnetic field. It can't be neglected in general in the problems of laser and nonlinear optics. It will be seen below that it is just this factor which would play an important role in the symmetry violation of time reversal in light's absorption and radiation processes.

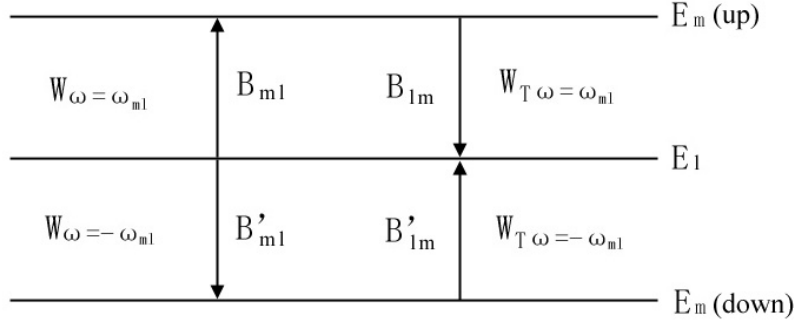

Fig.1. Electron's transitions among three energy levels

Let $\vec{R}_0$ represents the distance vector pointing from radiation source to atomic mass center, $\vec{r}$ represents the distance vector pointing from atomic mass center to electron, we have $\vec{R} = \vec{R}_0 + \vec{r}$. For the interaction process between external electromagnetic field and atom in medium, we have $R_0 = 0.1 \sim 1m$, $\vec{k} \cdot \vec{R}_0 = 10^6 \sim 10^7 \gg 1$ and $\vec{k} \cdot \vec{r} \ll 1$. If radiation fields come from atomic internal, we have $\vec{R}_0 \approx 0$, $\vec{k} \cdot \vec{r} \ll 1$. In the following discussion, we approximately take

$$e^{-i\vec{k} \cdot \vec{R}} \approx e^{-i\vec{k} \cdot \vec{R}_0}\left[1 - i\vec{k} \cdot \vec{r} - (\vec{k} \cdot \vec{r})^2 / 2\right] \tag{16}$$

In which $\vec{k} = \omega \vec{\tau} / c$, $\vec{\tau}$ is unit direction vector. By means of relations $\omega_{lm} = -\omega_{ml}$, $\hat{p} = -i\hbar \nabla$ and

$$\langle m|\hat{p}|l\rangle = \mu\langle m|d\hat{r}/dt|l\rangle = \frac{\mu}{i\hbar}\langle m|[\hat{r}, \hat{H}_0]|l\rangle = i\mu\omega_{ml}\langle m|\hat{r}|l\rangle \tag{17}$$

as well as Eq.(6), we can get

$$\hat{F}_{1ml} = \left[-\frac{iq\omega_{ml}}{2\omega}\vec{E}_0 \cdot \langle m|\vec{r}|l\rangle + \frac{q\hbar}{2c\mu}\vec{E}_0 \cdot \langle m|\vec{\tau} \cdot \vec{r}\nabla|l\rangle - \frac{iq\hbar\omega}{4c^2\mu}\vec{E}_0 \cdot \langle m|(\vec{\tau} \cdot \vec{r})^2\nabla|l\rangle\right]e^{-i\vec{k} \cdot \vec{R}_0} \tag{18}$$

$$\hat{F}_{1ml}^+ = \left[-\frac{iq\omega_{ml}}{2\omega}\vec{E}_0 \cdot \langle m|\vec{r}|l\rangle - \frac{q\hbar}{2c\mu}\vec{E}_0 \cdot \langle m|\vec{\tau} \cdot \vec{r}\nabla|l\rangle - \frac{iq\hbar\omega}{4c^2\mu}\vec{E}_0 \cdot \langle m|(\vec{\tau} \cdot \vec{r})^2\nabla|l\rangle\right]e^{i\vec{k} \cdot \vec{R}_0} \tag{19}$$

The first item is the result of dipolar moment interaction. The second item is the result of quadrupolar moment interaction and the third item is the result of octupolar moment interaction. The wave functions of stationary states $|m\rangle$ and $|l\rangle$ have fixed parities. The parities of operator $\vec{r}$ and $(\vec{\tau} \cdot \vec{r})^2\nabla$ are odd and the parity of operator $\vec{\tau} \cdot \vec{r}\nabla$ is even. So by the consideration of symmetry, if matrix element $\langle m|\vec{r}|l\rangle \neq 0$, we would have $\langle m|\vec{\tau} \cdot \vec{r}\nabla|l\rangle = 0$ and $\langle m|(\vec{\tau} \cdot \vec{r})^2\nabla|l\rangle \neq 0$. Conversely, if $\langle m|\vec{r}|l\rangle = 0$, we would have $\langle m|\vec{\tau} \cdot \vec{r}\nabla|l\rangle \neq 0$ and $\langle m|(\vec{\tau} \cdot \vec{r})^2\nabla|l\rangle = 0$. Suppose $\langle m|\vec{r}|l\rangle \neq 0$, $\langle m|\vec{\tau} \cdot \vec{r}\nabla|l\rangle = 0$ and $\langle m|(\vec{\tau} \cdot \vec{r})^2\nabla|l\rangle \neq 0$, we have $\hat{F}_{1ml} \neq \hat{F}_{1ml}^+$ but $|\hat{F}_{1ml}|^2 = |\hat{F}_{1ml}^+|^2$. So after retarded interaction is

taken into account for the first order processes, we still have $W^{(1)}_{\omega=\omega_{ml}} = W^{(1)}_{\omega=-\omega_{ml}}$, i.e., the transition probabilities of stimulated radiation and stimulated absorption are still the same.

## 3. The time reversal of the first order process

Let's discuss the time reversal of the first order process below. According to the standard theory of quantum electrodynamics, the time reversal of electromagnetic potential is $\vec{A}(\vec{x},t) \to -\vec{A}(\vec{x},-t)$. Meanwhile, we have $\hat{p} \to -\hat{p}$ when $t \to -t$. The propagation direction of electromagnetic wave should be changed from $\vec{k}$ to $-\vec{k}$ under time reversal (Otherwise retarded wave would become advanced wave so that the law of causality would be violated.). Let subscript $T$ represent time reversal, from Eqs.(3), (5) and (6), we have $\hat{H}'_{1T}(\vec{x},t) = \hat{H}'_1(\vec{x},t)$ and $\hat{H}'_{2T}(\vec{x},t) = \hat{H}'_2(\vec{x},t)$. The interaction Hamiltonian is unchanged under time reversal. On the other hand, when $t \to -t$, Eq. (8) becomes

$$-i\hbar \frac{\partial}{\partial t}|\psi\rangle_T = (\hat{H}_{0T} + \hat{H}'_T)|\psi\rangle_T \qquad |\psi\rangle_T = \sum_n a_n(-t) e^{\frac{i}{\hbar}E_n t}|n\rangle \qquad (20)$$

Let $a_m(-t) = a_m^{(0)}(-t) + a_m^{(1)}(-t) + a_m^{(2)}(-t) + \cdots$ and put it into the formula above, the motion equation becomes

$$-i\hbar \frac{d}{dt}\left[a_m^{(0)}(-t) + a_m^{(1)}(-t) + a_m^{(2)}(-t) + \cdots\right]$$
$$= \sum_n (H'_{1Tmn} + H'_{2Tmn})\left[a_n^{(0)}(-t) + a_n^{(1)}(-t) + a_n^{(2)}(-t) + \cdots\right]e^{-i\omega_{mn}t} \qquad (21)$$

Take index replacements $m \to l$ and $n \to k$ in the formula, then let the items with same order to be equal to each other, we get

$$-i\hbar \frac{d}{dt}a_l^{(0)}(-t) = 0 \qquad -i\hbar \frac{d}{dt}a_l^{(1)}(-t) = \sum_k \hat{H}'_{1Tlk} a_k^{(0)}(-t) e^{-i\omega_{lk}t} \qquad (22)$$

$$-i\hbar \frac{d}{dt}a_l^{(2)}(-t) = \sum_k \hat{H}'_{2Tlk} a_k^{(0)}(-t) e^{-i\omega_{lk}t} + \sum_k \hat{H}'_{1Tlk} a_k^{(1)}(-t) e^{-i\omega_{lk}t} \qquad (23)$$

$$-i\hbar \frac{d}{dt}a_l^{(3)}(-t) = \sum_k \hat{H}'_{2Tlk} a_k^{(1)}(-t) e^{-i\omega_{lk}t} + \sum_k \hat{H}'_{1Tlk} a_k^{(2)}(-t) e^{-i\omega_{lk}t} \qquad (24)$$

On the other hand, under time reversal, the initial state becomes $|m\rangle$ with $a_k^{(0)}(-t) = \delta_{km}$. Put it into the second formula of Eq.(22), we get

$$a_l^{(1)}(-t) = -\frac{1}{i\hbar}\int_0^t \hat{H}'_{1Tlm} e^{-i\omega_{lm}t} dt = -\frac{1}{i\hbar}\int_0^t \hat{H}'_{1lm} e^{-i\omega_{lm}t} dt \qquad (25)$$

Because $\hat{H}'_1$ is the Hermitian operator with $\hat{H}'_{1lm} = \langle l|\hat{H}'_1|m\rangle = \langle m|\hat{H}'_1|l\rangle^* = \hat{H}'^*_{1ml}$, we can write $\hat{H}'^*_{1ml} = \hat{F}'_{1ml} e^{i\omega t} + \hat{F}'^+_{1ml} e^{-i\omega t}$ and get

$$\hat{F}'_{1ml} = \left(\hat{F}^+_{1ml}\right)^* = \left[\frac{iq\omega_{ml}}{2\omega}\vec{E}_0\cdot\langle m|\vec{r}|l\rangle^* - \frac{q\hbar}{2c\mu}\vec{E}_0\cdot\langle m|\vec{\tau}\cdot\vec{r}\nabla|l\rangle^* + \frac{iq\hbar\omega}{4c^2\mu}\vec{E}_0\cdot\langle m|(\vec{\tau}\cdot\vec{r})^2\nabla|l\rangle^*\right]e^{-i\vec{k}\cdot\vec{R}_0} \quad (26)$$

$$\hat{F}'^+_{1ml} = \left(\hat{F}_{1ml}\right)^* = \left[\frac{iq\omega_{ml}}{2\omega}\vec{E}_0\cdot\langle m|\vec{r}|l\rangle^* + \frac{q\hbar}{2c\mu}\vec{E}_0\cdot\langle m|\vec{\tau}\cdot\vec{r}\nabla|l\rangle^* + \frac{iq\hbar\omega}{4c^2\mu}\vec{E}_0\cdot\langle m|(\vec{\tau}\cdot\vec{r})^2\nabla|l\rangle^*\right]e^{i\vec{k}\cdot\vec{R}_0} \quad (27)$$

Let $a_{mT}(t)$ represent the time reversal of amplitude $a_m(t)$. Because the original final state becomes into $|l\rangle$ and the original initial state becomes into $|m\rangle$ under time reversal, we have $a^{(1)}_{mT}(t) = a^{(1)}_l(-t)$. So according to Eq.(25), after time reversal, the transition amplitude becomes

$$a^{(1)}_{mT}(t) = -\frac{1}{i\hbar}\int_0^t \hat{H}'^*_{1ml}e^{i\omega_{ml}t}dt = \frac{\hat{F}'_{1ml}\left[e^{i(\omega+\omega_{ml})t}-1\right]}{\hbar(\omega+\omega_{ml})} - \frac{\hat{F}'^+_{1ml}\left[e^{-i(\omega-\omega_{ml})t}-1\right]}{\hbar(\omega-\omega_{ml})}$$

$$= \frac{\hat{F}'_{1ml}\left[e^{i(\omega-\omega_{lm})t}-1\right]}{\hbar(\omega-\omega_{lm})} - \frac{\hat{F}'^+_{1ml}\left[e^{-i(\omega+\omega_{lm})t}-1\right]}{\hbar(\omega+\omega_{lm})} \quad (28)$$

So the condition $\omega = -\omega_{lm} = \omega_{ml}$ corresponds to the situation with $E_m > E_l$, indincating that an electron emits a photon with energy $-\hbar\omega_{lm} = E_m - E_l > 0$ and transits from the initial high-energy state $|m\rangle$ into the final low-energy state $|l\rangle$. This process is the time reversal of stimulated absorption process described by Eq.(14). By considering Eq.(27), the transition probability in unite time is

$$W^{(1)}_{T\omega=\omega_{ml}} = \frac{2\pi}{\hbar^2}\left|\hat{F}'^+_{1ml}\right|^2\delta(\omega+\omega_{lm}) = \frac{2\pi}{\hbar^2}\left|\hat{F}_{1ml}\right|^2\delta(\omega-\omega_{ml}) \quad (29)$$

Comparing with Eq.(14) and considering the result $\left|\hat{F}_{1ml}\right|^2 = \left|\hat{F}^+_{1ml}\right|^2$, we still have $W^{(1)}_{T\omega=\omega_{ml}} = W^{(1)}_{\omega=\omega_{ml}}$. Because we define the time reversal of stimulated absorption process as the stimulated radiation process, the result shows that the transition probability of stimulated absorption is equal to that of stimulated radiation after time reversal for the first order process when retarded interaction is considered. The process is unchanged under time reversal.

Similarly, the condition $\omega = \omega_{lm} = -\omega_{ml}$ corresponds to the situation with $E_m < E_l$, indicating that an electron emits a photon with energy $\hbar\omega_{lm} = E_l - E_m > 0$ and transits from the initial low-energy state $|m\rangle$ into the final high-energy state $|l\rangle$. This process is the time reversal of stimulated radiation process described by Eq.(15). By considering Eq.(26), the transition probability in unite time is

$$W^{(1)}_{T\omega=-\omega_{ml}} = \frac{2\pi}{\hbar^2}\left|\hat{F}'_{1ml}\right|^2\delta(\omega-\omega_{lm}) = \frac{2\pi}{\hbar^2}\left|\hat{F}^+_{1ml}\right|^2\delta(\omega+\omega_{ml}) \quad (30)$$

After retarded effect is considered, we also have $W^{(1)}_{T\omega=-\omega_{ml}} = W^{(1)}_{\omega=-\omega_{ml}}$ for the first order process. The process is unchanged under time reversal.

The transition relations can be seen clearly in Fig.1. There are two stimulated absorption parameters $B_{ml}$, $B'_{lm}$ and two stimulated radiation parameters $B_{lm}$, $B'_{ml}$. The initial and final states of $B_{ml}$ and $B_{lm}$ are opposite. So they are time reversal states. The initial and final states of $B'_{ml}$ and $B'_{lm}$ are also opposite, so they are also time reversal states. Therefore, if $B_{ml}$ is defined as stimulated absorption parameter, $B_{lm}$ should be defined as stimulated radiation parameter for their initial final states are just opposite. Similarly, if $B'_{ml}$ is defined as stimulated absorption parameter, $B'_{lm}$ should be defined as stimulated radiation parameter. Meanwhile, if $B_{ml}$ (or $B'_{lm}$) is defined as stimulated absorption parameter, $B'_{ml}$ (or $B_{lm}$) should not be defined as stimulated radiation parameter, for they have same initial and final states and do not describe corresponding stimulated radiation and absorption processes. For the first order process, we have $B_{ml} = B_{lm} = B'_{lm} = B'_{ml}$. But as shown below that in the high order processes, this relation can't hold.

## 4. The transition probability of the second order process

The second order processes are discussed below. We write $\hat{H}'_2 = \hat{F}_2 e^{2i\omega t} + \hat{F}_2^+ e^{-2i\omega t} + \hat{F}_0$, in which

$$\hat{F}_2 = \frac{q^2 E_0^2}{2\omega^2 \mu} e^{-i2\vec{k}\cdot\vec{R}} \qquad \hat{F}_2^+ = \frac{q^2 E_0^2}{2\omega^2 \mu} e^{i2\vec{k}\cdot\vec{R}} \qquad \hat{F}_0 = \frac{q^2 E_0^2}{\omega^2 \mu} \qquad (31)$$

When $m \neq l$ we have $\langle m|l\rangle = 0$. Suppose $\langle m|\vec{\tau}\cdot\vec{r}|l\rangle \neq 0$, we have $\langle m|(\vec{\tau}\cdot\vec{r})^2|l\rangle = 0$. According to Eq.(16), we have

$$\hat{F}_{2ml} = -\frac{iq^2 E_0^2}{c\omega\mu} e^{-i2\vec{k}\cdot\vec{R}_0} \langle m|\vec{\tau}\cdot\vec{r}|l\rangle \qquad \hat{F}_{2ml}^+ = \frac{iq^2 E_0^2}{c\omega\mu} e^{i2\vec{k}\cdot\vec{R}_0} \langle m|\vec{\tau}\cdot\vec{r}|l\rangle \qquad \hat{F}_{0ml} = 0 \qquad (32)$$

Similarly, we suppose the initial condition is $a_n^{(0)} = \delta_{nl}$. Substituting Eq.(13) into Eq.(13) and taking the integral, we can obtain the transition probability amplitude of the second order process

$$a_m^{(2)}(t) = \frac{1}{i\hbar}\int_0^t \hat{H}'_{2ml} e^{i\omega_{ml}t} dt + \frac{1}{i\hbar}\sum_n \int_0^t \hat{H}'_{1mn} a_n^{(1)}(t) e^{i\omega_{mn}t} dt$$

$$= -\frac{\hat{F}_{2ml}\left[e^{i(2\omega+\omega_{ml})t}-1\right]}{\hbar(2\omega+\omega_{ml})} + \frac{\hat{F}_{2ml}^+\left[e^{-i(2\omega-\omega_{ml})t}-1\right]}{\hbar(2\omega-\omega_{ml})}$$

$$+\sum_n \frac{\hat{F}_{1mn}\hat{F}_{1nl}}{\hbar^2(\omega+\omega_{nl})}\left\{\frac{e^{i(2\omega+\omega_{nl}+\omega_{mn})t}-1}{2\omega+\omega_{nl}+\omega_{mn}} - \frac{e^{i(\omega+\omega_{mn})t}-1}{\omega+\omega_{mn}}\right\}$$

$$-\sum_n \frac{\hat{F}_{1mn}\hat{F}_{1nl}^+}{\hbar^2(\omega-\omega_{nl})}\left\{\frac{e^{i(\omega_{nl}+\omega_{mn})t}-1}{\omega_{nl}+\omega_{mn}} - \frac{e^{i(\omega+\omega_{mn})t}-1}{\omega+\omega_{mn}}\right\}$$

$$+\sum_n \frac{\hat{F}_{1mn}^+\hat{F}_{1nl}}{\hbar^2(\omega+\omega_{nl})}\left\{\frac{e^{i(\omega_{nl}+\omega_{mn})t}-1}{\omega_{nl}+\omega_{mn}} + \frac{e^{-i(\omega-\omega_{mn})t}-1}{\omega-\omega_{mn}}\right\}$$

$$+\sum_n \frac{\hat{F}_{1mn}^+\hat{F}_{1nl}^+}{\hbar^2(\omega-\omega_{nl})}\left\{\frac{e^{-i(2\omega-\omega_{nl}-\omega_{mn})t}-1}{2\omega-\omega_{nl}-\omega_{mn}} - \frac{e^{-i(\omega-\omega_{mn})t}-1}{\omega-\omega_{mn}}\right\} \qquad (33)$$

The formula contains the transition processes of single photon's absorption and radiation with $\omega = \pm\omega_{ml}$ as well as the processes of double photon's absorption and radiation with $2\omega = \pm\omega_{ml}$. We only discuss the absorption process of a single photon here. By rotation wave approximation, only the items containing factor $\{e^{-i(\omega-\omega_{ml})t}-1\}/(\omega-\omega_{ml})$ are remained. Let $n=l$ in the fifth and sixth items of the formula, the transition probability amplitude is

$$a_m^{(2)}(t)_{\omega=\omega_{ml}} = \frac{\hat{F}_{1ml}^+(\hat{F}_{1ll}-\hat{F}_{1ll}^+)[e^{-i(\omega-\omega_{ml})t}-1]}{\hbar^2\omega(\omega-\omega_{ml})} = \frac{\hat{F}_{1ml}^+(\hat{F}_{1ll}-\hat{F}_{1ll}^+)[e^{-i(\omega-\omega_{ml})t}-1]}{\hbar^2\omega_{ml}(\omega-\omega_{ml})} \quad (34)$$

Therefore, after the second order process is considered, the total transition probability amplitude is

$$a_m(t)_{\omega=\omega_{ml}} = a_m^{(1)}(t)_{\omega=\omega_{ml}} + a_m^{(2)}(t)_{\omega=\omega_{ml}} = \frac{\hat{F}_{1ml}^+[e^{-i(\omega-\omega_{ml})t}-1]}{\hbar(\omega-\omega_{ml})}\left(1+\frac{\hat{F}_{1ll}-\hat{F}_{1ll}^+}{\hbar\omega_{ml}}\right) \quad (35)$$

Because we have $\omega_{ll}=0$ and $\langle l|\vec{r}|l\rangle=0$, when $m=l$, the first item in Eqs.(18) and (19) are equal to zero, but the second items are not equal to zero in general. By the consideration of symmetry, the third item are also zero, or can be neglected by comparing with the second item. So it is enough for us only to consider quadrupolar moment interaction in this case. We have

$$\hat{F}_{1ll}-\hat{F}_{1ll}^+ = \frac{q\hbar}{c\mu}\vec{E}_0\cdot\langle l|\vec{\tau}\cdot\vec{r}\nabla|l\rangle\cos\vec{k}\cdot\vec{R}_0 = B_{1l}+iB_{2l} \quad (36)$$

Let

$$A_l^2 = B_{1l}^2 + 2\hbar\omega_{ml}B_{1l} + B_{2l}^2 \quad (37)$$

When $\omega=\omega_{ml}$ we get the transition probability of the second order stimulated absorption process

$$W_{\omega=\omega_{ml}}^{(2)} = \frac{2\pi}{\hbar^2}|\hat{F}_{1ml}|^2\left\{1+\frac{A_l^2}{\hbar^2\omega_{ml}^2}\right\}\delta(\omega-\omega_{ml}) \quad (38)$$

The magnitude order of the revised value of the second order process is estimated below. The wave function of bounding state's atoms can be developed into series with form $|l\rangle \sim \sum b_n(\theta,\varphi)r^n$ in general. If $\langle l|\vec{\tau}\cdot\vec{r}\nabla|l\rangle\neq 0$, we have $\vec{\tau}\cdot\vec{r}\nabla|l\rangle \sim r\partial/(\partial r)|l\rangle \sim |l\rangle$ approximately, or $\langle l|\vec{\tau}\cdot\vec{r}\nabla|l\rangle \sim \langle l|l\rangle \sim 1$. By taking $\omega_{ml}=10^{16}$, we have

$$\frac{A_l^2}{\hbar\omega_{ml}^2} \sim 4q^2E_0^2|\langle l|\vec{\tau}\cdot\vec{r}\nabla|l\rangle|^2/(c\mu\omega_{ml})^2 \sim 1.4\times 10^{-26}E_0^2 \quad (39)$$

In weak electromagnetic fields with $E_0 << 10^{13}V/m$, the revised values of the second processes can be neglected. When the fields are strong enough with $E_0 \approx 10^{12} \sim 10^{13}V/m$, the revised value is big enough to be observed. The revised factor $A_l$ of the second process is only relative to initial state, having nothing to do with final states. On the other hand, if the retarded effect of radiation fields is neglected with $\vec{k}\cdot\vec{r} \sim \vec{\tau}\cdot\vec{r}=0$, we have $\langle l|\vec{\tau}\cdot\vec{r}\nabla|l\rangle=0$, the revised value of the second order process vanishes.

## 5. The time reversal of the second order process

The time reversal of the second order process is discussed now. Under time reversal, the initial state becomes $a_k^{(0)}(-t) = \delta_{km}$. By means of relations $\hat{H}'_{1Tlm} = \hat{H}'^*_{1ml}$ and $\hat{H}'_{2Tlm} = \hat{H}'^*_{2ml}$, we get the time reversal of transition amplitude for the second order processes according to Eq.(11)

$$a_l^{(2)}(-t) = -\frac{1}{i\hbar}\int_0^t \hat{H}'_{2Tlm} e^{-i\omega_{lm}t} dt - \frac{1}{i\hbar}\sum_k \int_0^t \hat{H}'_{1Tlk} a_k^{(1)}(-t) e^{-i\omega_{lk}t} dt$$

$$= -\frac{1}{i\hbar}\int_0^t \hat{H}'^*_{2ml} e^{i\omega_{ml}t} dt - \frac{1}{i\hbar}\sum_k \int_0^t \hat{H}'^*_{1kl} a_k^{(1)}(-t) e^{i\omega_{kl}t} dt \tag{40}$$

Let $\hat{H}'^*_{2ml} = \hat{F}'_{2ml} e^{i2\omega t} + \hat{F}'^+_{2ml} e^{-i2\omega t} + \hat{F}'_{0ml}$, when $m \neq l$, we have $\langle m|l\rangle = 0$ and

$$\hat{F}'_{2ml} = -\frac{iq^2 E_0^2}{c\omega\mu} e^{-i2\vec{k}\cdot\vec{R}_0} \langle m|\vec{\tau}\cdot\vec{r}|l\rangle^* \qquad \hat{F}'^+_{2ml} = \frac{iq^2 E_0^2}{c\omega\mu} e^{i2\vec{k}\cdot\vec{R}_0} \langle m|\vec{\tau}\cdot\vec{r}|l\rangle^* \qquad \hat{F}'_{0ml} = 0 \tag{41}$$

So the time reversal of transition probability amplitude of the second process is

$$a_{Tm}^{(2)}(t) = a_l^{(2)}(-t) = \frac{\hat{F}'_{2ml}\left[e^{i(2\omega+\omega_{ml})t} - 1\right]}{\hbar(2\omega+\omega_{ml})} - \frac{\hat{F}'^+_{2ml}\left[e^{-i(2\omega-\omega_{ml})t} - 1\right]}{\hbar(2\omega-\omega_{ml})}$$

$$+ \sum_k \frac{\hat{F}'^+_{1mk}\hat{F}'^+_{1kl}}{\hbar^2(\omega-\omega_{mk})}\left\{\frac{e^{-i(2\omega-\omega_{mk}-\omega_{kl})t} - 1}{2\omega-\omega_{mk}-\omega_{kl}} - \frac{e^{-i(\omega-\omega_{kl})t} - 1}{\omega-\omega_{kl}}\right\}$$

$$- \sum_k \frac{\hat{F}'^+_{1mk}\hat{F}'_{1kl}}{\hbar^2(\omega-\omega_{mk})}\left\{\frac{e^{i(\omega_{mk}+\omega_{kl})t} - 1}{\omega_{mk}+\omega_{kl}} - \frac{e^{i(\omega+\omega_{kl})t} - 1}{\omega+\omega_{kl}}\right\}$$

$$+ \sum_k \frac{\hat{F}'_{1mk}\hat{F}'^+_{1kl}}{\hbar^2(\omega+\omega_{mk})}\left\{\frac{e^{i(\omega_{mk}+\omega_{kl})t} - 1}{\omega_{mk}+\omega_{kl}} + \frac{e^{-i(\omega-\omega_{kl})t} - 1}{\omega-\omega_{kl}}\right\}$$

$$+ \sum_k \frac{\hat{F}'_{1mk}\hat{F}'_{1kl}}{\hbar^2(\omega+\omega_{mk})}\left\{\frac{e^{i(2\omega+\omega_{mk}+\omega_{kl})t} - 1}{2\omega+\omega_{mk}+\omega_{kl}} - \frac{e^{i(\omega+\omega_{kl})t} - 1}{\omega+\omega_{kl}}\right\} \tag{42}$$

When $\omega = \omega_{ml}$, we take $k = m$ in the third and fifth items of the formula. Also by rotation wave approximation, the time reversal of probability amplitude is

$$a_{mT}^{(2)}(t)\Big|_{\omega=\omega_{ml}} = \frac{\hat{F}'^+_{1ml}(F'_{1mm} - F'^+_{1mm})\left[e^{-i(\omega-\omega_{ml})t} - 1\right]}{\hbar^2\omega(\omega-\omega_{ml})} = \frac{\hat{F}'^+_{1ml}(F'_{1mm} - F'^+_{1mm})\left[e^{-i(\omega-\omega_{ml})t} - 1\right]}{\hbar^2\omega_{ml}(\omega-\omega_{ml})} \tag{43}$$

The time reversal of the total stimulated absolution process is

$$a_{Tm}(t)\Big|_{\omega=\omega_{ml}} = a_{Tm}^{(1)}(t)\Big|_{\omega=\omega_{ml}} + a_{Tm}^{(2)}(t)\Big|_{\omega=\omega_{ml}} = \frac{\hat{F}'^+_{1ml}\left[e^{-i(\omega-\omega_{ml})t} - 1\right]}{\hbar(\omega-\omega_{ml})}\left(1 + \frac{\hat{F}'_{1mm} - \hat{F}'^+_{1mm}}{\hbar\omega_{ml}}\right) \tag{44}$$

Comparing with Eq.(34), because of $\hat{F}'_{mm} \neq \hat{F}_{ll}$, we have $a_{mT}(t)\Big|_{\omega=\omega_{ml}} \neq a_m(t)\Big|_{\omega=\omega_{ml}}$, the transition

probability amplitude can not keep unchanged. Similarly, by considering $\omega_{mm} = 0$ and from Eqs.(26) and (27), we have

$$\hat{F}'_{1mm} - \hat{F}'^{+}_{1mm} = -\frac{q\hbar}{c\mu}\vec{E}_0 \cdot \langle m|\vec{\tau} \cdot \vec{r}\nabla|m\rangle^* \cos\vec{k} \cdot \vec{R}_0 = -B_{1m} + iB_{2m} \tag{45}$$

Let

$$A'^2_m = B^2_{1m} - 2\hbar\omega_{ml}B_{1m} + B^2_{2m} \tag{46}$$

When $\omega = \omega_{ml}$, the time reversal of stimulated absorption probability of the second process is

$$W^{(2)}_{T\omega=\omega_{ml}} = \frac{2\pi}{\hbar^2}\left|\hat{F}'^{+}_{1ml}\right|^2\left\{1 + \frac{A'^2_m}{\hbar^2\omega^2_{ml}}\right\}\delta(\omega - \omega_{ml}) = \frac{2\pi}{\hbar^2}\left|\hat{F}_{1ml}\right|^2\left\{1 + \frac{A'^2_m}{\hbar^2\omega^2_{ml}}\right\}\delta(\omega - \omega_{ml}) \tag{47}$$

The revised factor $A_m$ is also only relative to initial state. Because $A'_m \neq A_l$, we have

$$W^{(2)}_{\omega=\omega_{ml}}\left(\hat{F}_{1ml}, A_l\right) \neq W^{(2)}_{T\omega=\omega_{ml}}\left(\hat{F}_{1ml}, A'_m\right) \tag{48}$$

The second process of stimulated absolution violates time reversal symmetry. The parameter of symmetry violation of the second order process can be is defined as

$$\beta = \frac{W^{(2)}_{T\omega=\omega_{ml}} - W^{(2)}_{\omega=\omega_{ml}}}{W^{(2)}_{\omega=\omega_{ml}}} \sim \frac{(A^2_m - A^2_l)}{\hbar^2\omega^2_{ml}} \sim 10^{-26} E^2_0 \tag{49}$$

When the radiation fields are strong enough with $E_0 \approx 10^{12} \sim 10^{13} V/m$, the time reversal symmetry violation of the second order process would be great.

Meanwhile, by means of Eqs.(33) and (42), for the second process with $\omega = -\omega_{ml}$, the transition amplitude and probability can be obtained with

$$a^{(2)}_m(t)_{\omega=-\omega_{ml}} = \frac{\hat{F}_{1ml}\left(\hat{F}_{1ll} - \hat{F}^{+}_{1ll}\right)\left[e^{i(\omega+\omega_{ml})t} - 1\right]}{\hbar^2\omega_{ml}(\omega + \omega_{ml})} \tag{50}$$

$$W^{(2)}_{\omega=-\omega_{ml}} = \frac{2\pi}{\hbar^2}\left|\hat{F}_{1ml}\right|^2\left\{1 + \frac{A^2_l}{\hbar^2\omega^2_{ml}}\right\}\delta(\omega + \omega_{ml}) \tag{51}$$

Their time reversals are

$$a^{(2)}_m(t)_{T\omega=-\omega_{ml}} = \frac{\hat{F}'_{1ml}\left(\hat{F}'_{1mm} - \hat{F}'^{+}_{1mm}\right)\left[e^{i(\omega+\omega_{ml})t} - 1\right]}{\hbar^2\omega_{ml}(\omega + \omega_{ml})} \tag{52}$$

$$W^{(2)}_{T\omega=-\omega_{ml}} = \frac{2\pi}{\hbar^2}\left|\hat{F}_{1ml}\right|^2\left\{1 + \frac{A''^2_m}{\hbar^2\omega^2_{ml}}\right\}\delta(\omega + \omega_{ml}) \tag{53}$$

Also, the process violates time reversal symmetry. It is easy to prove that for the second order processes of double photon absorptions with $2\omega = \pm\omega_{ml}$, the transition probabilities are unchanged under time reversal. The symmetry violation would appear in the third order processes. This problem will be discussed in detail later.

Let $B_{ml}$ represent the stimulated absorption probability of an electron (in unit radiation density and unit time) transiting from initial low-energy state $|l\rangle$ to final high-energy state $|m\rangle$, $B_{lm}$ represent the probability of stimulated radiation of an electron (in unit radiation density and unit time) transiting from the initial high-energy state $|m\rangle$ into the final low-energy state to $|l\rangle$, we have

$$B_{ml} = \frac{4\pi^2}{3\hbar^2}|\vec{D}_{ml}|^2(1+\chi_{ml}) \qquad B_{lm} = (B_{ml})_T = \frac{4\pi^2}{3\hbar^2}|\vec{D}_{ml}|^2(1+\chi'_{ml}) \qquad (54)$$

Here $\vec{D}_{ml}$ is the dipolar moment of an electron. We have $\chi_{ml} \neq \chi'_{ml}$ and $B_{ml} \neq B_{lm}$ in general, i.e., the parameters of light's stimulated radiation and absolution are not the same. It also means that the nonlinear optical processes would violate time reversal symmetry in general. This problem will be discussed in detail later.

## 6. Accumulate solution of double energy level system and time reversal

What discussed above is that the radiation fields are polarized and monochromatic light. It is easy to prove that when the radiation fields are non-polarized and non-monochromatic light, time reversal symmetry is still violated after the retarded effect of radiation field is taken into account in the high order processes. But we do not discuss this problem here. The approximate method of perturbation is used in the discussion above. In order to prove that symmetry violation of time reversal is not introduced by the approximate method, we discuss double energy level system below. The wave function of double energy lever system can be written as

$$|\psi\rangle = a(t)e^{-\frac{i}{\hbar}E_1 t}|1\rangle + b(t)e^{-\frac{i}{\hbar}E_2 t}|2\rangle \qquad (55)$$

Thus the motion equations of quantum mechanics are

$$i\hbar \dot{a}(t) = \hat{H}'_{11}a(t) + \hat{H}'_{12}e^{-i\omega_{21}t}b(t) \qquad i\hbar \dot{b}(t) = \hat{H}'_{21}e^{i\omega_{21}t}a(t) + \hat{H}'_{22}b(t) \qquad (56)$$

For simplification, we only consider the first item of the Hamiltonian Eq.(2) to take $\hat{H}'_1 \neq 0$, $\hat{H}'_2 = 0$ and $\hat{H}' = \hat{H}'_1$. By taking dipolar approximation with $\vec{k} \cdot \vec{R} = 0$, we have $\langle 1|\hat{H}'|1\rangle = \langle 2|\hat{H}'|2\rangle = 0$. By the rotation wave approximation again, the motion equations becomes

$$\ddot{a}(t) - i(\omega - \omega_{21})\dot{a}(t) + A^2 a(t) = 0 \qquad \ddot{b}(t) + i(\omega - \omega_{21})\dot{b}(t) + A^2 b(t) = 0 \qquad (57)$$

Here $A = |\hat{F}_{21}|^2/\hbar^2$. These two equations have accurate solutions, i.e., the so-called Rabi solutions. Suppose that atom is in the state $|1\rangle$ at beginning with $a(t=0)=1$ and $b(t=0)=0$, we can get [1]

$$|b(t)|^2 = \frac{4A^2 \sin^2\sqrt{(\omega-\omega_{21})^2 + 4A^2}\, t/2}{(\omega-\omega_{21})^2 + 4A^2} \qquad (58)$$

If the atom is in the state $|2\rangle$ at beginning with $b(t=0)=1$ and $a(t=0)=0$, we have

$$|a(t)|^2 = \frac{4A^2 \sin^2 \sqrt{(\omega - \omega_{21})^2 + 4A^2}\, t/2}{(\omega - \omega_{21})^2 + 4A^2} \tag{59}$$

So for the Rabi process, the probability that atom transits from $|1\rangle$ into $|2\rangle$ is the same as that atom transits from $|2\rangle$ into $|1\rangle$. The processes are symmetrical under time reversal. In fact, let $t \to -t$ in Eq.(57), we get

$$\ddot{a}(-t) + i(\omega - \omega_{21})\dot{a}(-t) + V^2 a(-t) = 0 \qquad \ddot{b}(-t) - i(\omega - \omega_{21})\dot{b}(-t) + V^2 b(-t) = 0 \tag{60}$$

Comparing these two formulas with Eq.(57), we know that as long as let $a(-t) = b(t)$ and $b(-t) = a(t)$, the motion equations are the same under time reversal.

However, if retarded effect is considered with $\vec{k} \cdot \vec{R} \neq 0$, we have $\langle 1|\hat{H}'|1\rangle \neq 0$ and $\langle 2|\hat{H}'|2\rangle \neq 0$. By taking $\hat{H}'_1 \neq 0$, $\hat{H}'_2 = 0$ similarly, we have

$$\ddot{a}(t) - i\left\{\frac{1}{\hat{H}'_{12}}\left[(\omega - \omega_{21})\hat{F}_{12}e^{i\omega t} - (\omega + \omega_{21})\hat{F}_{12}e^{-i\omega t}\right] - \frac{1}{\hbar}\left(\hat{H}'_{11} + \hat{H}'_{22}\right)\right\}\dot{a}(t)$$

$$- \left\{\frac{\hat{H}'_{11}}{\hbar \hat{H}'_{12}}\left[(\omega - \omega_{21})\hat{F}_{12}e^{i\omega t} - \frac{1}{\hbar}(\omega + \omega_{21})\hat{F}_{12}e^{-i\omega t}\right]\right.$$

$$\left. + \frac{\omega}{\hbar}\left(\hat{F}_{12}e^{i\omega t} - \hat{F}^+_{12}e^{-i\omega t}\right) + \frac{1}{\hbar^2}\left(\hat{H}'_{11}\hat{H}'_{22} - \hat{H}'_{12}\hat{H}'_{21}\right)\right\}a(t) = 0 \tag{61}$$

$$\ddot{b}(t) + i\left\{\frac{1}{\hat{H}'_{12}}\left[(\omega - \omega_{21})\hat{F}^+_{12}e^{-i\omega t} - (\omega + \omega_{21})\hat{F}_{12}e^{i\omega t}\right] - \left(\hat{H}'_{11} + \hat{H}'_{22}\right)\right\}\dot{b}(t)$$

$$- \left\{\frac{\hat{H}'_{22}}{\hbar \hat{H}'_{12}}\left[(\omega - \omega_{21})\hat{F}^+_{12}e^{-i\omega t} - \frac{1}{\hbar}(\omega + \omega_{21})\hat{F}_{12}e^{i\omega t}\right]\right.$$

$$\left. + \frac{\omega}{\hbar}\left(\hat{F}_{12}e^{i\omega t} - \hat{F}^+_{12}e^{-i\omega t}\right) + \frac{1}{\hbar^2}\left(\hat{H}'_{11}\hat{H}'_{22} - \hat{H}'_{12}\hat{H}'_{21}\right)\right\}b(t) = 0 \tag{62}$$

The equations have no accurate solutions in this case. Under time reversal, we have $t \to -t$, $\hat{H}_{T12} = \hat{H}'^*_{21}$, $\hat{F}_{T12} = \hat{F}'_{21}$, $\hat{F}^+_{T12} = \hat{F}'^+_{21}$, so the formulas above become

$$\ddot{a}(-t) + i\left\{\frac{1}{\hat{H}'^*_{21}}\left[(\omega - \omega_{21})\hat{F}'_{21}e^{-i\omega t} - (\omega + \omega_{21})\hat{F}'_{21}e^{i\omega t}\right] - \frac{1}{\hbar}\left(\hat{H}'^*_{11} + \hat{H}'^*_{22}\right)\right\}\dot{a}(-t)$$

$$- \left\{\frac{\hat{H}'^*_{11}}{\hbar \hat{H}'^*_{21}}\left[(\omega - \omega_{21})\hat{F}'_{21}e^{-i\omega t} - \frac{1}{\hbar}(\omega + \omega_{21})\hat{F}^*_{21}e^{i\omega t}\right]\right.$$

$$\left. + \frac{\omega}{\hbar}\left(\hat{F}'_{21}e^{-i\omega t} - \hat{F}'^+_{21}e^{i\omega t}\right) + \frac{1}{\hbar^2}\left(\hat{H}'^*_{11}\hat{H}'^*_{22} - \hat{H}'^*_{21}\hat{H}'^*_{21}\right)\right\}a(-t) = 0 \tag{63}$$

$$\ddot{b}(-t) - i\left\{\frac{1}{\hat{H}_{21}^{\prime*}}\left[(\omega-\omega_{21})\hat{F}_{21}^{\prime+}e^{i\omega t} - (\omega+\omega_{21})\hat{F}_{21}^{\prime}e^{-i\omega t}\right] - (\hat{H}_{11}^{\prime*} + \hat{H}_{22}^{\prime*})\right\}\dot{b}(-t)$$

$$-\left\{\frac{\hat{H}_{22}^{\prime*}}{\hbar\hat{H}_{21}^{\prime*}}\left[(\omega-\omega_{21})\hat{F}_{21}^{\prime+}e^{i\omega t} - \frac{1}{\hbar}(\omega+\omega_{21})\hat{F}_{21}^{\prime}e^{-i\omega t}\right]\right.$$

$$\left.+\frac{\omega}{\hbar}\left(\hat{F}_{21}^{\prime}e^{-i\omega t} - \hat{F}_{21}^{\prime+}e^{i\omega t}\right) + \frac{1}{\hbar^2}\left(\hat{H}_{11}^{\prime*}\hat{H}_{22}^{\prime*} - \hat{H}_{21}^{\prime*}\hat{H}_{12}^{\prime*}\right)\right\}b(-t) = 0 \quad (64)$$

Because of $\hat{H}_{11}^{\prime} \neq \hat{H}_{22}^{\prime*}$, $\hat{F}_{21}^{\prime} \neq \hat{F}_{21}^{\prime+}$ and $\hat{F}_{21}^{\prime} \neq \hat{F}_{21}^{\prime*}$, even by taking $a(-t) \to b(t)$, $b(-t) \to a(t)$, the motion equations can't yet keep unchanged under time reversal. So after retarded effect of radiation field is considered, the double energy level system can't keep unchanged under time reversal. It means that symmetry violation of time reversal is an inherent character of systems, not originates from the approximate method of perturbation.

## 7. Influence on the Foundational Theory of Laser

The influence of this paper's revision on the foundational theory of laser is discussed below. Let us first to discuss the system of double energy levels. Let $N_1$ represent the number of electrons on low energy level and $N_2$ represent the number of electrons on high energy level. Because of $B_{12} \neq B_{21}$ after the revision, under the condition $N_2 < N_1$ without the reversion of electron's population, as long as $B_{21}$ is big enough than $B_{12}$, we still have $B_{21}N_2 > B_{12}N_1$. That is to say, laser can be caused without the reversion of electron's population. At present, a great of numbers experiments verify this result [6], though the explanations are not the same. In light of this paper, this is a normal result. In fact, the electron's numbers on a energy level can not be determined by the experiments at present. What can be determined by experiments is the number of photons. And the numbers of photons are calculated by $\rho B_{21}N_2$, $\rho B_{12}N_1$ and $A_{21}N_2$. By considering the high order retarded effect of radiation fields, the condition of stimulated amplification to cause laser should be changed into $B_{21}N_2 > B_{12}N_1$, in stead of $N_2 > N_1$.

Secondly, according to the current theory, we have at least three energy levels to produce laser. For the systems with two energy levels, there is a so-called final balance with $B_{12}N_1\rho(\nu) = A_{21}N_2 + B_{21}N_2\rho(\nu)$. If $B_{12} = B_{21}$ and $A_{21} = k_{21}B_{21}$, we have

$$\frac{N_2}{N_1} = \frac{\rho}{(\rho + \kappa_{21})} < 1 \quad (65)$$

The result would be $N_2 < N_1$. In this case, we have no population reversion so that no laser is caused. According this paper, suppose we still have $A_{21} = k_{21}B_{21}$, when the balance is reached, we still have

$$\frac{N_2}{N_1} = \frac{\rho B_{12}}{(\rho + \kappa_{21})B_{21}} \quad (66)$$

Because of $B_{12} \neq B_{21}$, as long as relation $\rho B_{12} > (\rho + \kappa_{21})B_{21}$ is satisfied, we still have $N_2 > N_1$ so that population reversion can still be caused. But in this case, we have $B_{21}N_2 < B_{12}N_1$. That is to say, for the steady system of double energy levels, even though population reversion appeals, laser can not yet be caused. For the non-steady system of double energy levels, we have two cases

$$\frac{dN_2}{dt} = B_{12}N_1\rho - A_{21}N_2 - B_{21}N_2\rho > 0 \tag{67}$$

$$\frac{dN_2}{dt} = B_{12}N_1\rho - A_{21}N_2 - B_{21}N_2\rho < 0 \tag{68}$$

When $dN_2/dt > 0$, we have $B_{12}N_1\rho > A_{21}N_2 + B_{21}N_2\rho$, so $B_{12}N_1\rho > B_{21}N_2\rho$, no laser is produced. When $dN_2/dt < 0$, we have $B_{12}N_1\rho < A_{21}N_2 + B_{21}N_2\rho$. In this case, if $B_{12}N_1\rho < B_{21}N_2\rho$, laser can be produced. If $B_{12}N_1\rho > B_{21}N_2\rho$, laser can be produced.

Then we discuss the influence on the system of three energy levels. The stander stimulated radiation and absorption processes in the system of three energy levels is shown in Fig.7. In the current theory, however, the processes to produce laser is actually simplified as shown in Fig8. By analyzing the difference between them, we can know the significance of this paper's revision. According to Fig.8, when particles which are located on ground state $E_1$ at beginning are pumped into $E_3$ energy level, they can transit into $E_2$ energy level through both ways radiation transition and non-radiation transition. The population reversion can be achieved between $E_1$ and $E_2$, so that the laser with frequency $\omega_{21}$ can be produced. Comparing with Fig.7, the process shown in Fig.8 omits the spontaneous radiation and stimulated radiation transitions, as well as particle's transition from $E_2$ level into $E_3$ level. According to the Einstein's theory, we have $B_{13} = B_{31}$. The impossibilities are the same for a particle transiting from ground state into $E_3$ energy level and transiting from $E_3$ energy level back into ground state. Suppose that the number of particles which transit from ground state into $E_3$ energy level is $\rho(v_{13})B_{13}N_1$ in unit time. Thus there would have $\rho B_{31}N_3$ particles transiting back into ground state from $E_3$ energy level by stimulated radiation, and $A_{31}N_3 = \kappa_{31}B_{31}N_3$ particles transiting back into ground state from $E_3$ energy level by spontaneous radiation. Therefore, most of particles which have transited into $E_3$ energy level would come back into original ground state by emitting photons with frequency $\omega_{31}$, so that population reversion between $E_1$ and $E_2$ energy levels would be affected greatly. Meanwhile, because of $B_{23} = B_{32}$, some particles on $E_2$ energy level which came from $E_3$ energy level would transit back into $E_3$ energy level by stimulated absorption, so that population reversion between $E_1$ and $E_2$ would also be decreased. These results indicate that the Einstein's theory is only suitable for equilibrium processes actually, but the production of laser is equilibrium process.

The current theory of laser uses a hazy method to avoid theses problems. The probability a particle transiting back into ground state from $E_3$ energy level is not considered directly. In stead, we use a pumping speed $R$ replaces $\rho_{13}B_{31}N_1 - (\rho_{13} + \kappa_{31})B_{31}N_3$. On the other hand, the non-radiation transition is used to replace $(\rho_{23} + \kappa_{32})B_{32}N_3 - \rho_{23}B_{32}N_2$. In this way, the complexity of process is simplified.

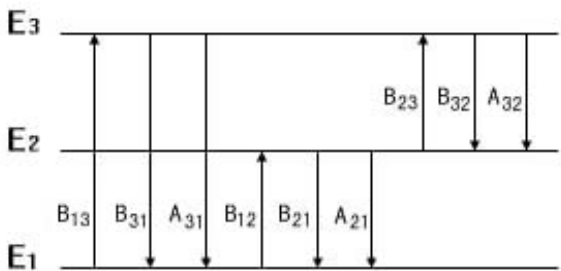 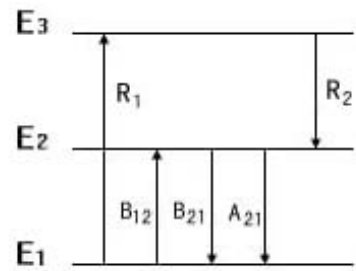

**Fig.7 Transition among three energy levels**    **Fig.8 Simplification of Fig.7**

According to the revision of is paper, we have $B_{ml} \neq B_{lm}$. Based on it, we can provide a more simple and rational picture for the production of laser in the system of three energy levels. In this case, we can have $B_{31} \ll B_{13}$ and $(\rho_{13} + \kappa_{31})B_{31}N_3 \ll \rho_{13}B_{31}N_1$, so that only a few particles can transit back into ground state from $E_3$ energy level by stimulated radiation and spontaneous radiation after they transit from ground state into $E_3$ energy level. Most of particles on $E_3$ energy level would transit into $E_2$ energy level. On the other hand, because of $B_{23} \ll B_{32}$, we have $\rho_{23}B_{23}N_2 \ll (\rho_{23} + \kappa_{32})B_{32}N_3$. Most of particles can not transit back into $E_3$ energy level after thy transit into $E_2$ energy from $E_3$ energy level. Meanwhile, because of $B_{12} \ll B_{21}$, it is difficult for particles on ground state to transit into $E_2$ energy level from ground state by stimulated absorption, but is easy to transit into ground state from $E_2$ energy level by stimulated radiation. Therefore, a high effective and ideal laser system of three energy levels should satisfy conditions $B_{21} \ll B_{12}$, $B_{31} \ll B_{13}$ and $B_{23} \ll B_{32}$. It is obvious that as long as we think $B_{ml} \neq B_{lm}$, we can more simply and rationally explain the produce of laser of the system of three energy levels.

In this way, we can also explain the phenomenon of optical self-transparence and self absorptions well[7]. Experiments show that in strong electric fields, some medium would have the saturated absorption of light, so that the medium would become transparent for light. The current explanation of saturated absorption is that the number $N_1$ of particles located on low energy level becomes small so that the stimulated absorption would becomes small for the absorption of light direct ratio to the number of particles located on low energy level. Meanwhile, the transmission light would be increased due to the stimulated radiation of particles located on high energy level, i.e., the self-transparence phenomena of saturated absorption appears. The problem of this explanation is that if the number $N_1$ of particles located on low energy level decreases and the number $N_2$ of particles located on high energy level increase, the spontaneous radiation would also increase. When stationary states are reached, we always have $A_{21}N_2$ photons emitted in the form of spontaneous radiation in unit time. Because spontaneous radiation takes place in all directions in space, it is difficult for medium to achieve real transparence. According the revised theory of this paper, we have revised factor $\alpha_{ml} \sim E_0^2$. If $\alpha_{ml} < 0$ so that $\alpha_{ml} \sim -1$ for some medium in strong field, the absorption parameter $B_{ml}$ would become very small even with $B_{ml} \sim 0$. In this case, even though a great number of particles are still located on low energy level, the saturated absorption of light would is caused so that the medium become transparence. In light of the current theory, we have $B_{ml} \sim E_0^2$. When $E_0$ increases, the absorption parameter would increase so that it is impossible for us to have $B_{ml} \sim 0$. Conversely, if $\alpha_{ml} > 0$ with $\alpha_{ml} \sim E_0^2$, light's absorption for some mediums would increase greatly in strong field. This is just the phenomena of self absorption. In the current non-linear optics, the phenomena of self absorption are explained the absorptions of double photons or multi-photons, as well as stimulated scattering. In light of this paper, besides the absorptions of double photons or multi-photons, the process of single photon would also cause trans-normal absorption. It is obvious that the revised theory can explain theses phenomena more rationally.

## 8. Discussion on the reasons of symmetry violation of time reversal

We need to discuss the reason of the symmetry violation of time reversal In the paper, semi-classical method is used, i.e., quantum mechanics is used to describe charged particles and classical electromagnetic theory is used to describe radiation fields. The limitation of this method is that spontaneous radiation can not be deduced automatically from the theory. The spontaneous radiation formula has to be obtained indirectly by means of the Einstein's theory of light's radiation and absorption. In strictly, we should discuss the problems using complete quantum theory, from which we can deduce spontaneous radiation probability automatically. However, as we known, except the spontaneous radiation, the results are the complete same by using both the semi-classical method and the complete quantum method to calculate light's stimulated radiation and absorption probabilities. It also means that if we use complete quantum mechanics to discuss light's stimulated radiation and absorption, time reversal symmetry would be also violated after the retarded effects of radiation fields are taken into account. It is just the spontaneous radiation which indicates the asymmetry of time reversal in the processes of interaction between light and charged particles, for there exits only light's spontaneous radiation without light's spontaneous absolution in nature. This result is completely asymmetrical. In fact, in complete quantum mechanics, we only let the factor $-q\vec{E}_0/2\omega\mu$ in the interaction Hamiltonian in semi-classical theory correspond to photon's creation or annihilation operators $\hat{a}^+$ or $\hat{a}$. This kind of correspondence does not change the results of time reversal symmetry violation in calculation processes. The problem is that if photon's creation or annihilation operators are used, some complexness and troubles would be caused in high order processes so that it may be too difficult to calculate. So in the problems of light's stimulated radiation and absorption and nonlinear optics, we use actually semi-classical even complete classical theory and methods and always obtain satisfied results at present.

Because the interaction Hamiltonian and the motion equation of quantum mechanics are unchanged under time reversal, what causes the symmetry violation of time reversal? The method of rotation wave approximation is used in the paper, if does this approximation method cause the symmetry violation of time reversal? Let's discuss this problem below.

Suppose that micro-states are described by $|\psi\rangle$ and $|\varphi\rangle$. Their time reversal are $|\psi_T\rangle = T|\psi\rangle$ and $|\varphi_T\rangle = T|\varphi\rangle$. Suppose that the interaction Hamiltonian is unchanged under time reversal, according to quantum mechanics, we have the so-called detail balance formula

$$\langle\psi|\hat{H}|\varphi\rangle = \langle\varphi_T|\hat{H}|\psi_T\rangle^* \tag{69}$$

It indicates that the transition probability is unchanged under time reversal in the quantum transition process. For the problem of light's stimulated radiation and absolution, if the radiation field is only a single frequency one, the interaction Hamiltonian is

$$\hat{H} = \hat{F}_0 + \hat{F}_1 e^{i\omega t} + \hat{F}_1^+ e^{-i\omega t} + \hat{F}_2 e^{i2\omega t} + \hat{F}_1^+ e^{-i2\omega t} \tag{70}$$

Meanwhile, if only a single particle state is considered, we have

$$|\psi\rangle = |\varphi\rangle = \sum_m a_m(t) e^{-\frac{i}{\hbar}E_m t}|m\rangle \qquad |\psi_T\rangle = |\varphi_T\rangle = \sum_m a_m(-t) e^{\frac{i}{\hbar}E_m t}|m\rangle \tag{71}$$

Substitute Eqs.(70) and (71) in to (69), we obtain

$$\sum_{m,l} e^{\frac{i}{\hbar}(E_m-E_l)t} a_m^*(t) a_l(t) \langle m|\hat{F}_0 + \hat{F}_1 e^{i\omega t} + \hat{F}_1^+ e^{-i\omega t} + \hat{F}_2 e^{i2\omega t} + \hat{F}_1^+ e^{-i2\omega t}|l\rangle$$

$$= \sum_{m,l} e^{-\frac{i}{\hbar}(E_m-E_l)t} a_m^*(-t) a_l(-t) \langle l|\hat{F}_0 + \hat{F}_1 e^{i\omega t} + \hat{F}_1^+ e^{-i\omega t} + \hat{F}_2 e^{i2\omega t} + \hat{F}_1^+ e^{-i2\omega t}|m\rangle^* \tag{72}$$

This is a result of multinomial sums. It means that the total transition probability is unchanged under time reversal. However, by the restriction of energy conservation law, in the formula above, only a few items which satisfy the condition $E_m - E_l = \pm n\hbar\omega$ can be actualized really. Those items which do not satisfy the condition are forbidden actually. Remaining the items which satisfy the condition of energy conservation and giving up the items which do not, the procedure is just the so-called rotation wave approximation. It is obvious that the two sides of Eq.(68) would not be equal to each other again after the procedure is carried out, i.e., the symmetry of time reversal would be violated.

The paper calculates the transition and time reversal problems of partial items corresponding to the operators $\hat{F}_1 e^{-i\omega t}$ and $\hat{F}_1^+ e^{i\omega t}$ under the situation $n=1$. Because Eq.(68) can not be accurately calculated, we use approximate method to let $a_m(t) = a_m^{(0)}(t) + a_m^{(1)}(t) + a_m^{(2)}(t) + \cdots$. For the first order approximation, we have $a_m(t) = a_m^{(0)}(t) + a_m^{(1)}(t)$. Suppose that an atom transits from state $\langle l|$ into state $\langle m|$, we get transition probability Eq.(14) and its time reversal Eq.(29). It indicates that the first process is unchanged under time reversal after retarded effect of radiation field is considered. For the second processes, let $a_m(t) = a_m^{(0)}(t) + a_m^{(1)}(t) + a_m^{(2)}(t)$ and suppose in the same way that an atom transits from state $\langle l|$ into state $\langle m|$, we get Eq.(38) and its time reversal Eq.(47). The result violates the symmetry of time reversal and the symmetry violation is relative to the asymmetry of initial states of bounding state atoms before and after time reversal. The uniform values of the Hamiltonian operator about the initial states of an atom before and after time reversal are not equal to each other. Therefore, one of the reasons to cause the symmetry violation of time reversal is that the condition of energy conservation forbids some transition processes between bounding state atoms, so that realizable processes violates time reversal symmetry with

$$W_{\omega=\pm\omega_{ml}}(\hat{F}_{1ml}) + W_{2\omega=\pm\omega_{ml}}(\hat{F}_{1ml}, A_l) + W_{3\omega=\pm\omega_{ml}}(\hat{F}_{1ml}, A_l) + \cdots$$

$$\neq W_{T\omega=\pm\omega_{ml}}(\hat{F}_{1ml}) + W_{T2\omega=\pm\omega_{ml}}(\hat{F}_{1ml}, A'_m) + W_{T3\omega=\pm\omega_{ml}}(\hat{F}_{1ml}, A'_m) + \cdots \quad (73)$$

Meanwhile, for concrete atoms, the other restriction conditions just as the wave function's symmetries and so on should be also considered. So only a few and specific transitions can be achieved actually. Most processes shown in Eq.(73) can not be completed. These realizable processes are just what we can observe and measure. They are irreversible in general. Therefore, the symmetry violation of time reversal in the filial or partial processes of light's stimulated radiation and absolution do not contradict with the detail balance formula (69) actually.

On the other hand, it can be seen from the discussion above that in the first order process of stimulated absolution, only operator $\hat{F}_1^+$ acts. In the second order process of stimulated absolution, besides the main action of operator $\hat{F}_1^+$, the operator $\hat{F}_1$ also affects the revised value. But in the first order process of stimulated radiation, only operator $\hat{F}_1$ acts. In the second order process of stimulated absolution, besides the main action of operator $\hat{F}_1$, the operator $\hat{F}_1^+$ also affects the revised value. This is to say, the actions of effective transition operators are also asymmetric under time reversal. This is also one of the reasons to cause the symmetry violation of time reversal. In the same time, the symmetry violation of time reversal is also relative to the asymmetry of initial states of bound atoms before and after time reversal. For the interaction between the radiation fields and the unbound atoms with continuous energy levels, there exists no symmetry violation of time reversal, for there exists no the asymmetry problem of the initial states before and after time reversal. Meanwhile, there is the difference of a negative sign between $A_l$ shown in Eq. (37) and $A'_m$ shown in Eq. (46). This difference is caused by the interference of amplitudes between the first order and the second order processes before and after time reversal. But if the retarded effect of radiation field is neglected, the processes of light's stimulated radiation and absolution would be symmetrical under time reversal. So the reasons of symmetry violation of time reversal are caused by multi-factors and quite complex.

The significance of this result is that it can provide us a method to solve a great problem of so-called reversibility paradox which has puzzled physical circle for along time. According to present understanding, micro-processes are considered reversible under time reversal, but macro-processes controlled by the second law of thermodynamics are always irreversible. We do not know how to solve this contradiction up to now$^{(8)}$. Though many theories have been advanced, for example, the theories of coarseness and mixing current and so on$^{(9)}$, none is satisfied. As we known that macro-systems are composed of atoms and molecules, and atoms and molecules are composed of charged particles. By the photon's radiations and absorptions, the interactions between the charged particles of bounding states and radiation fields take place. According to discussion in the paper, after the retarded effects of radiation fields are considered, the time reversal symmetry of light's stimulated radiation and absolution would be violated, even though the interaction Hamiltonian is unchanged under time reversal. Only when the systems reach macro-equilibrium states, or the probabilities of micro-particles radiating and absorbing photons are the same from the angle of statistical average, the processes are reversible under time reversal. Therefore, it can be said that irreversibility of macro-processes originates from the irreversibility of micro-processes actually.

The author gratefully acknowledges valuable discussions with Professors Qiu Yishen in Physical and Optical Technology College, Fujian Normal University and Zheng Shibiao in Physical Department, Fuzhou University.